\documentclass[runningheads]{llncs}
\usepackage[colorlinks, citecolor=blue]{hyperref}
\usepackage{booktabs} 
\usepackage{setspace, amsmath, amssymb, url, lscape, subfig, algorithmicx, multirow, pslatex, listings, verbatim, alltt, amsfonts, wrapfig, boxedminipage, color, bookmark, pifont, algpseudocode}
\usepackage{caption} 
\usepackage[vlined,linesnumbered,ruled,boxed]{algorithm2e}
\usepackage{epsfig}
\usepackage{xcolor}
\usepackage{balance}
\usepackage{enumitem}

\newcommand{\eg}{{\it e.g., }}

\newcommand{\ie}{{\it i.e., }}

\newcommand{\comments}[1]{}
\newcommand\hl{\bgroup\markoverwith
  {\textcolor{yellow}{\rule[-.5ex]{2pt}{2.5ex}}}\ULon}
\algrenewcommand\algorithmicrequire{\textbf{Input:}}
\algrenewcommand\algorithmicensure{\textbf{Output:}}
\newlength{\boxfigwidth}
\newcommand{\boxfig}[1]{
\begin{figure}[h]
\begin{center}
\begin{small}
\setlength{\boxfigwidth}{4.48in}
\addtolength{\boxfigwidth}{0in}
\noindent\framebox{\quad\begin{minipage}{\boxfigwidth}
#1
\vspace{-15pt}
\end{minipage}\quad}
\end{small}
\end{center}
\end{figure}
}

\begin{document}
\title{Leveraging Computational Reuse for Cost- and QoS-Efficient Task Scheduling in Clouds}
%
%
\author{Chavit Denninnart \inst{1} \and
Mohsen Amini Salehi\inst{1} \and
Adel Nadjaran Toosi\inst{2} \and
Xiangbo Li\inst{3}
}\authorrunning{C. Denninnart et al.}

%
\institute{ School of Computing and Informatics,\\ University of Louisiana at Lafayette, Louisiana, USA 
\email{\{cxd9974,amini\}@louisiana.edu} \and
Faculty of Information Technology, Monash University, Victoria, Australia 
\email{adel.n.toosi@monash.edu} \and
Brightcove Inc. , Arizona, USA \email{xli@brightcove.com} }
\maketitle              
\begin{abstract}

Cloud-based computing systems could get oversubscribed due to budget constraints of cloud users which causes violation of Quality of Experience (QoE) metrics such as tasks' deadlines. We investigate an approach to achieve robustness against uncertain task arrival and oversubscription through smart reuse of computation while similar tasks are waiting for execution. Our motivation in this study is a cloud-based video streaming engine that processes video streaming tasks in an on-demand manner. We propose a mechanism to identify various types of ``mergeable'' tasks and determine when it is appropriate to aggregate tasks without affecting QoS of other tasks. 
Experiment shows that our mechanism can improve robustness of the system  and also saves the overall time of using cloud services by more than 14\%.

%
%
\end{abstract}




%
%
%

\section{Introduction}
\label{sec:intro}
With Cloud and Edge Computing gaining more popularity as the back-end platform of many applications
, the need for efficient use of these platforms is of paramount importance for individual users and businesses.
A common practice to efficiently utilize cloud resources is to use a central queue of arriving tasks with a scheduler that allocates these tasks to a scalable pool of worker Virtual Machines (VMs). The tasks often have individual deadlines that failure to meet them compromises the Quality of Experience (QoE) expected by the end-users. 

Although cloud providers supply virtually unlimited resources, users generally have budget constraints, thus, cannot lavishly acquire cloud resources (VMs)~\cite{bi2017application}. Such limitation raises the \emph{oversubscription} problem, particularly, when there is a surge in the tasks arriving to the system. An oversubscribed system 
is defined as a system that is overwhelmed with arriving tasks to the extent that there is no way to meet the deadlines of all the tasks, thus, violating end-users' QoE. 

A large body of research has been dedicated to alleviate the oversubscription problem. The approaches undertaken in these research works follow two main lines of thinking; First, \emph{allocation-based approaches} 
 that try to minimize the impact of oversubscription through efficient mapping (scheduling) of the tasks to the resources. Second, approaches based on \emph{computational reuse} 
 that avoid or alleviates the oversubscription through efficient caching of the computational results. 

Although both of the aforementioned approaches are effective, they are limited in certain ways. The allocation-based approaches cannot entirely resolve the oversubscription because there is no such a solution according to the above-mentioned definition. In addition, many of the approaches are based on complex scheduling algorithms that impose extra overhead to the already oversubscribed system. 
The approaches based on computational reuse are also limited because they can only reuse the computations for tasks that are identical to the ones already completed and cached. In other words, if two tasks share part of their computation, it cannot be captured by current caching techniques. 

In this research, we propose a mechanism based on computational reuse that aims at alleviating oversubscription by aggregating identical and similar tasks  in the scheduling queue. 
Our mechanism makes the scheduling queue less busy and potentially lighten up the overhead of the scheduling process. It complements the existing scheduling-based and caching-based approaches but is not a replacement for them.

We define \emph{mergeable} tasks as those tasks that are either identical or sharing part of their operation with other tasks. We need a mechanism to, first, detect different types of mergeable tasks and, second, eliminate the detected mergeable tasks from the scheduling queue without causing further deadline violations in the system. 


Our motivational case study in this research is a video streaming engine that needs to process videos (\eg downsizing resolution or bit-rate) in the cloud before streaming them to viewers~\cite{ahmad2005video}. In this system, it is likely that viewers request same videos to be streamed, hence, creating similar tasks in the system especially when the system is oversubscribed. For example, two viewers who use similar display devices may request to stream the same video with the same or different specifications. The former case creates \emph{identical} tasks in the system whereas the latter one creates \emph{similar} tasks.  

In this research, we develop an Admission Control component 
that is able to detect different levels of similarity between tasks. The system is aware of the tasks' deadlines and performs merging without introducing additional deadline violations. 
The task aggregation also results in efficient utilization of resources and enable more tasks to meet their deadlines. Therefore, both viewers and system providers can be benefited from the proposed mechanism. 
In summary, the \textbf{key contributions} of this research are as follows:
\textbf{(A)} Proposing an efficient way of identifying potentially mergeable tasks; \textbf{(B)} Determining appropriateness and potential side-effects of merging tasks; \textbf{(C)} Analyzing the performance of the task aggregation mechanism on the viewers' QoE and time of deploying cloud resources (VMs).

Although we develop this mechanism in the context of video streaming, the idea of task aggregation and research findings of this work are valid for other domains. 


\section{Background for Merge-Aware Admission Control}
\label{sec:arch}

While storing multiple versions of the same video to serve different types of display devices is a conventional practice, Cloud-based Video Streaming Engine (CVSE)~\cite{CVSSJournal} enables on-demand (\ie lazy) processing of video streams, particularly for rarely accessed video streams~\cite{darwich16}. 

\begin{wrapfigure}{r}{0.6\textwidth}
\includegraphics[width=0.62\textwidth ]{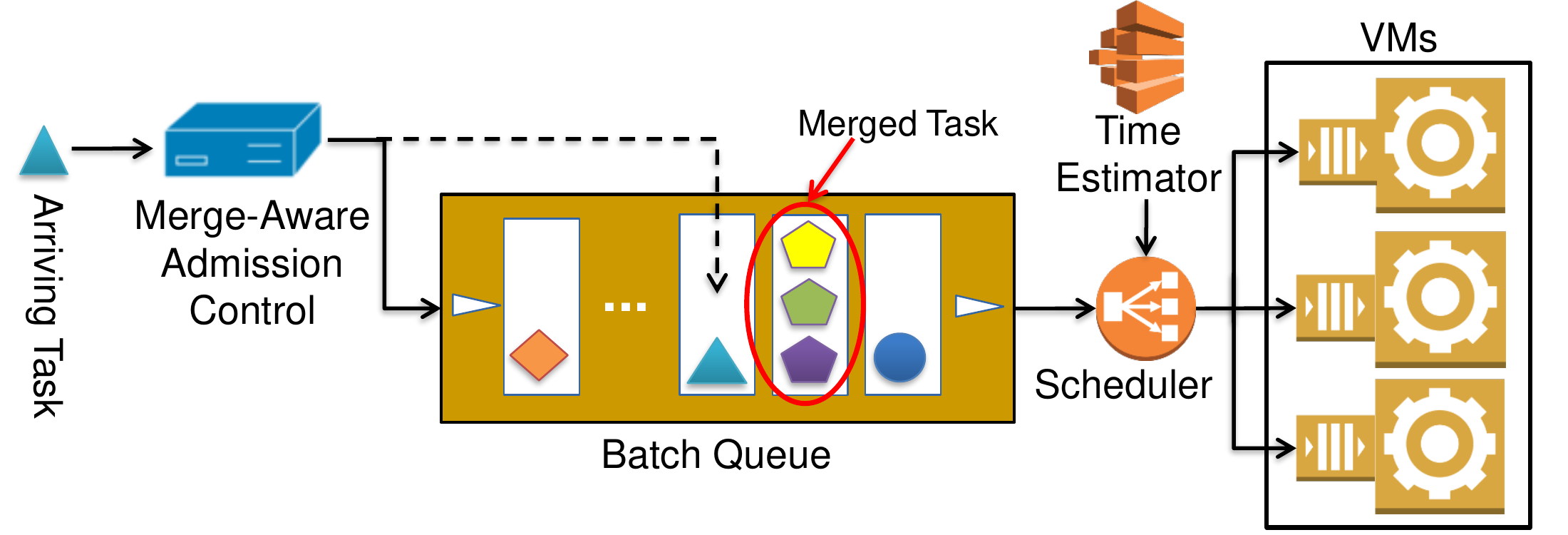}
    \caption{\small{Overview of Merge-Aware Admission Control for CVSE system}}
\label{fig:arch} 
\end{wrapfigure}

In the CVSE architecture, each task is a GOP (Group Of Picture) of the requested video stream. A GOP task request (hereafter, called task) includes the operation required along with the corresponding parameters bound to that request. 
Admission Control component, as shown in Figure~\ref{fig:arch}, sends the task to the batch queue (aka scheduling queue) where the task waits to be assigned by the scheduling policy~\cite{matin_paper} to one of multiple VMs' queues. Most of the scheduling policies are reliant on the Time Estimator component that is aware of the expected execution time of each task type (\eg different transcoding operations) on the cloud VMs~\cite{CVSS}. Tasks get executed on the assigned VM and streamed to the viewer (More details about CVSE is in \cite{CVSS,CVSSJournal}).


In this paper, we develop our task aggregation mechanism inside Admission Control component of CVSE. For an arriving task, Admission Control recognizes if it is mergeable with the ones exist in the batch queue or local queues of the VMs. Then, the Admission Control decides the feasibility of merging (\ie if merging causes deadline violation for other tasks). 

\section{Task Similarity Detection}

\label{sec:detection}

\subsection{Categories of Mergeable Tasks}
Mergeability of two given tasks can be explained based on the amount of computation the two tasks share. In particular, mergeability of two or more tasks can be achieved in the following levels: 
\begin{enumerate}[label=(\Alph*)]
 \item \emph{Task-level}: This is when more than one instance of the same task exists in the scheduling queue. Therefore, this level is also known as 
 \emph{Identical tasks} and can achieve maximum computational reusability. 
As these tasks are identical, merging them consumes the same resources required for only one task, hence, reducing both cost and processing delay.
 
 \item \emph{Operation-level}: This is when two or more tasks perform the same operation on the same data but with different configurations. In this level of merging, the two tasks can share part of their processing. 
 
 \item \emph{Data-level}: This is when two or more tasks perform different operations on the same data. 
This level of merging achieves the minimum reusability by saving only the time and processing overhead of loading data. 
\end{enumerate}

It is noteworthy that 
the aforementioned reusability levels are generic and can be further categorized depending on the context. 


\subsection{Detecting Similar Tasks}\label{sec:mergable-detection}

In this section, we provide a method to detect similar tasks. Although our solution carry out task aggregation using Admission Control component. We would like to note that, it is theoretically possible to carry out task merging in the scheduling queue, \ie after the task admission. In this case, to find mergeable tasks, we need to scan the entire queue and perform a pair-wise matching between the queued tasks. Practically, this approach is not efficient, because each time the queue is scanned, it imposes a significant number of redundant comparisons. Hence, we choose to perform task merging upon task arrival using the Admission Control component of the system.

Assuming there are $n$ tasks in the queue, for each arriving task, a na\"{i}ve mergeable task detection method 
has the overhead of performing $n$ comparisons to find the mergeable tasks. To reduce the overhead, we propose a method that functions based on the 
hashing techniques. The general idea of the proposed method is to generate a hash key from the arriving task request signature (\eg GOP id, processing type, and their parameters). Then, the Admission Control finds mergeable tasks by searching for a matching key in the hash table of tasks exist in the scheduling queue. 
%
%

The explained method can detect task-level mergeability. We need to expand it to detect operation- and data-level of task mergeabilities. To maximize the computational reusability, an arriving task is first verified against task-level mergeability. If there is no match in the task-level, then the method proceeds with checking the next levels of mergeability, namely operation-level and data-level, respectively. To achieve the multiple levels of mergeability, we create three hash-tables ---each covers one level of mergeability. The hash-keys in each level are constructed from the tasks' characteristics that are relevant in deciding mergeability at that level. For instance, in video streaming case study, keys in the hash-table that verifies task-level mergeability are constructed from GOP id, processing type, and their parameters. While, keys in the hash-table that verifies operation-level mergeability are constructed from GOP id and processing type. Similarly, keys in the hash-table of data-level mergeability are constructed from GOP id. 

\boxfig{
Upon arrival of task $j$: 
\begin{itemize}

\item[(1)] if $j$ merges with existing task $i$ on Task-level similarity:
\begin{itemize}
\item[--] No update on hash-table is required
\end{itemize}
\item[(2)] if $j$ merges with existing task $i$ on Operation- or Data-level similarity:
\begin{itemize}
\item[--] Add an entry to each hash-table with hash-keys of task $j$ and point them to merged task $i+j$ 
\end{itemize}
\item[(3)] if $j$ matches with existing task $i$ but the system chooses not to merge them:
\begin{itemize}
\item[--] Add an entry to each hash-table with hash-keys of task $j$ and point them to task $j$
\end{itemize}
\item[(4)] if $j$ does not match with any of the existing tasks:
\begin{itemize}
\item[--] Hash-keys of task $j$ are added to the respective hash-tables 
\end{itemize}
\end{itemize}

Upon task $j$ completing execution (\ie dequeuing task $j$):
\begin{itemize}
\item[--] Remove all entries pointing to task $j$ from hash-tables
\end{itemize}

\caption{The procedure to update hash-tables upon arrival or completion of tasks} \label{fig:htupdate}

}

Each entry of the hash-tables includes a hash-key and a pointer to the corresponding task. Entries of the three hash-tables must be updated upon a task arrival and execution. The only exception is Task-level merging, which does not require updating the hash-tables. Figure \ref{fig:htupdate} shows the procedure for updating the hash-tables for a given task $j$. 

When the system merges task $j$ with existing task $i$, the merged task, denoted as $i+j$, is essentially the object of task $i$ that is augmented with request information (\eg processing parameters) of task $j$. In this case, as shown in Step (2) of this procedure, the system only adds an entry to each hash-table with hash-key of task $j$ pointing to merged task $i+j$ as existing key for task $i$ already pointed to task $i+j$. 
When task $j$ is mergeable with existing task $i$, but the system decides to add task $j$ to the batch queue without merging. In this case, task $j$ has a higher likelihood of merging with other arriving tasks. 
 The reason is that task $i$ has not merged with task $j$ and it does not merge with other arriving tasks.
Hence, as shown in Step (3) of the procedure, the matching entry pointing to task $i$ is redirected and points to task $j$. 
It is worth noting that if the arriving task does not match with any of the existing tasks, as shown in Step (4), its hash-keys must be generated and added to the respective hash-tables. Also, when a task completes its execution, its corresponding entries are removed from the hash-tables. 


\section{Identifying Merging Appropriateness}
\label{sec:appropriateness}
Imagine an arriving task merges into an existing task in the queue. If such merging is not a Task-level similarity, the execution time of merged task is increased compared to a task before merging. The increased execution time delays the execution of other tasks waiting behind in the queue which could result in deadline violations. Therefore it is critical to assess the impact of merging tasks before performing the merge.

Impact of merging can be assess based on additional deadline misses of tasks following merged tasks when merging occurred against without. Impact of merging assessor create virtual copies of scheduling queue in two scenarios: with merging occurred and without. It simulates the scheduling and estimates completion time of each task, then compares to its deadline. Merging is only carried out if it does not cause additional deadline violations than it would normally happen if the tasks are not merged.

The estimated completion time of task $i$ on a given machine $m$, denoted as $C_i^m$ and formally shown in Equation~\ref{eq:compl}, is calculated as the sum of the four following factors: (A) current time, denoted $\tau$; (B) estimated remaining time to complete the currently executing task on machine $m$, denoted $e_r^m$; (C) sum of the estimated execution times of $N$ tasks pending in machine queue $m$, ahead of task $i$. This is calculated as $\sum_{p=1}^{N}(\mu_p + 2\cdotp \sigma_p)$; (D) estimated execution time of task $i$. 
 \begin{equation}\label{eq:compl}
  C_i^m = \tau + e_r^m + \sum_{p=1}^{N} (\mu_p +2\cdotp \sigma_p)  + (\mu_i + 2\cdotp \sigma_i)
 \end{equation} 

\section{Performance Evaluation}
\label{sec:evltn}
\vspace{-10pt}
\subsection{Experimental Setup}
We implemented a prototype of CVSE with task aggregation mechanism equipped. It is designed to operate in different modes, namely real streaming mode and emulation mode that is used for testing purposes~\cite{CVSSJournal}. 
In this study, to examine various workloads, we used CVSE in the emulation mode. We evaluated the proposed mechanism using eight homogeneous VMs modeled after Amazon GPU (\texttt{g2.2xlarge}) VM.  

The video repository we used for evaluation includes multiple replicas of a set of benchmark videos. Videos in the benchmarking set are diverse both in terms of the content types and length. The length of the videos in the benchmark varies in the range of [10, 600] seconds splitting into 10-110 Group Of Picture (GOP) segments. The benchmark videos are publicly available for reproducibility purposes at \url{https://goo.gl/TE5iJ5}.
For each GOP of the benchmark videos, we obtained their processing times by executing each processing operation 30 times on Amazon GPU VM. The processing operations we benchmarked are: reducing resolution, changing codec, adjusting bit rate, and changing frame rate. 

To evaluate the system under various workload intensities, we generate [2000, 3000] GOP processing tasks within a fixed time interval. We collect the deadline miss-rate (DMR) and makespan (\ie execution time to finish all tasks) of completing all tasks. We conducted each experiment 30 times, each time with random task arrival time and order. Mean and 95\% confidence interval of the results are reported. 
We examined three queuing policies, namely FCFS (First-Come-First-Serve), EDF (Earliest Deadline First), and MU (Max Urgency). For each queuing policy, we studied no task merging versus task merging. In the experiments, all tasks must be comepleted, even if they miss their deadline.

\subsection{Impact of Task Aggregation}
\begin{wrapfigure}{r}{0.40\textwidth}
	\centering
	\includegraphics[width=4.9cm]{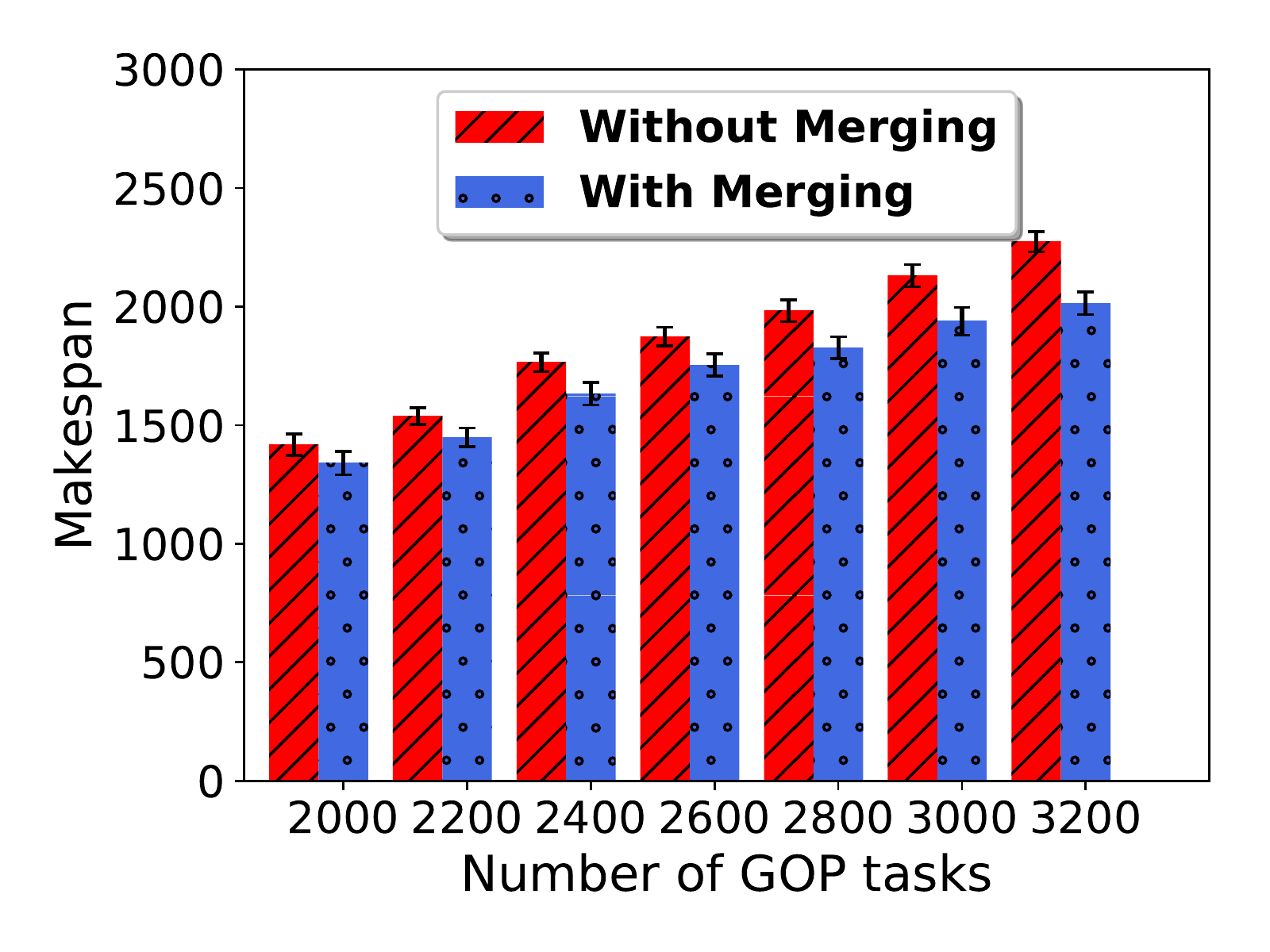} 
	\caption{\small{Comparing the total time to complete tasks (\ie makespan) under varying number of arriving GOP tasks (horizontal axes) in two scenarios: without task merging, and with task merging. }}%
	\label{fig:exetime}%
	
\vspace{-10pt}    
\end{wrapfigure}
\paragraph*{Evaluating Makespan:}
In the first experiment, our goal is to see the impact of task merging on makespan. In fact, makespan implies the time cloud resources are deployed, which implies the cost incurred to execute all the tasks. We examine the system under various subscription levels (from 2000 to 3200 GOPs) arriving within the same time interval. 
As we can see in Figure~\ref{fig:exetime}, our proposed merging mechanism saves between 4.40\% and 14.33\% in makespan. Execution time saving is more pronounced when the system is more oversubscribed. It is worth noting that makespan does not vary under different scheduling policies.

\begin{figure*}%
    \centering 
    \subfloat[\small{DMR under FCFS Queue}]{{\includegraphics[width=3.95cm]{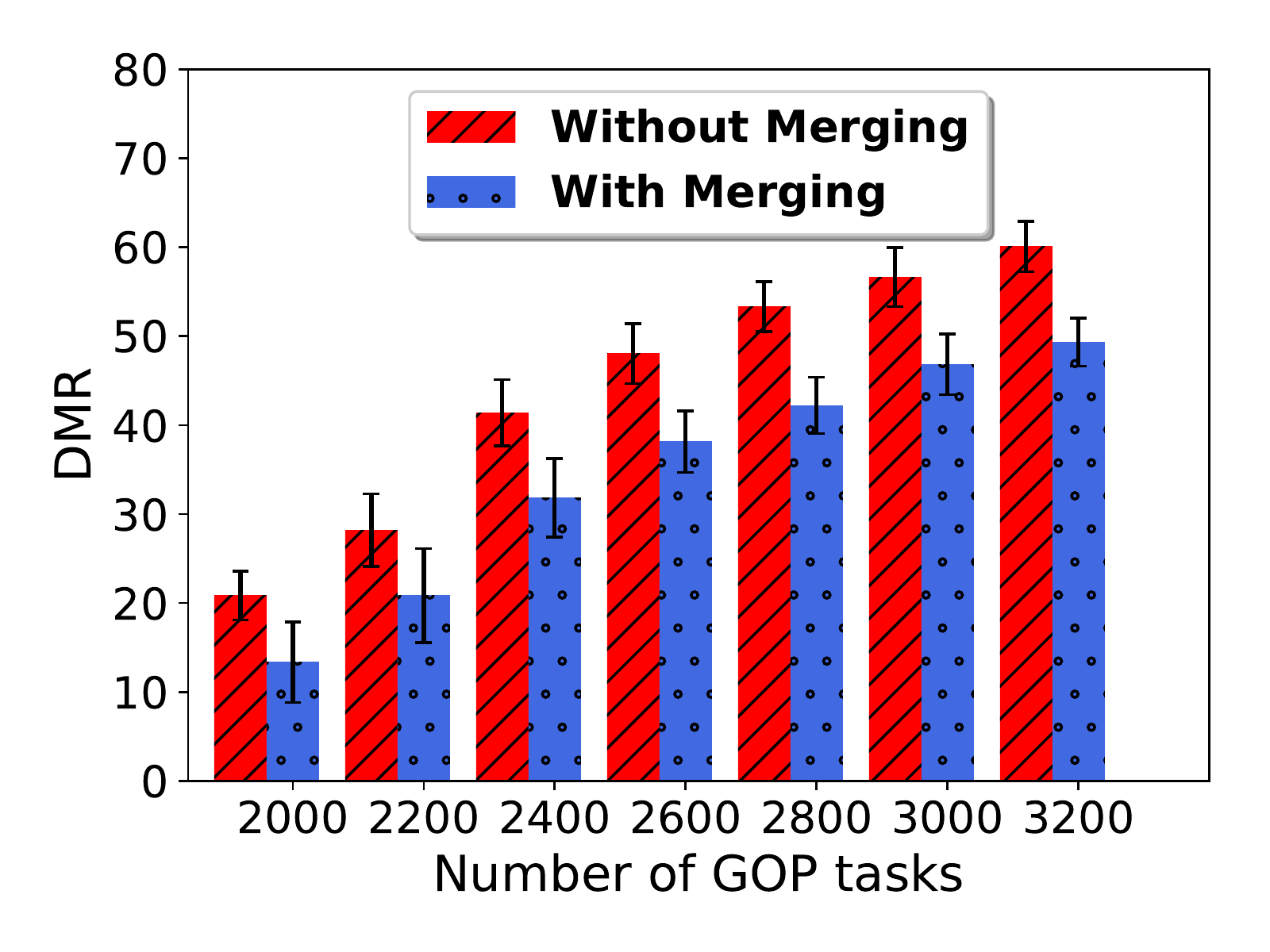} }}\hspace{0.03cm}%
    \subfloat[\small{DMR under EDF Queue}]{{\includegraphics[width=3.95cm]{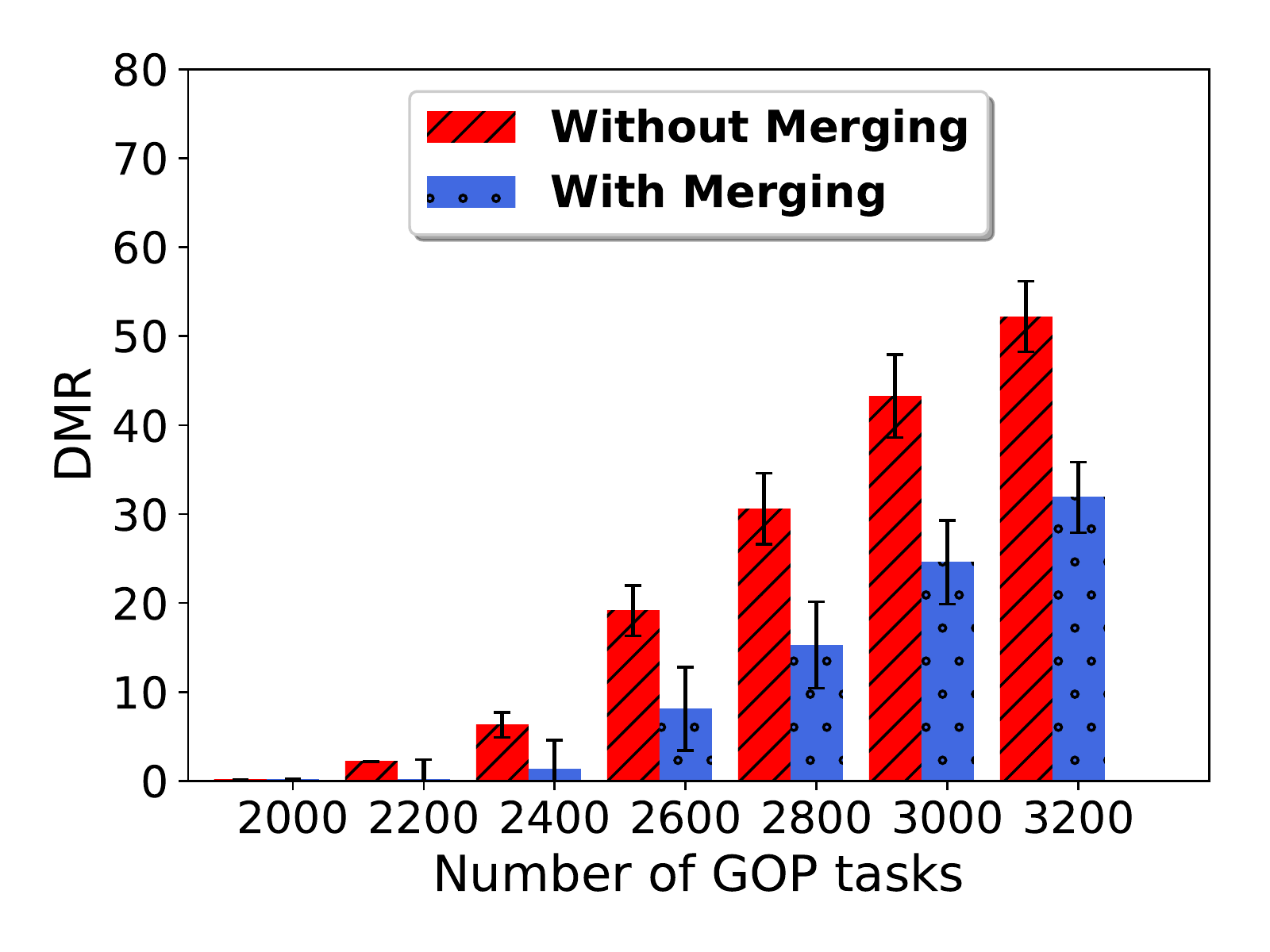} }}\hspace{0.03cm}%
    \subfloat[\small{DMR under MU Queue}]{{\includegraphics[width=3.95cm]{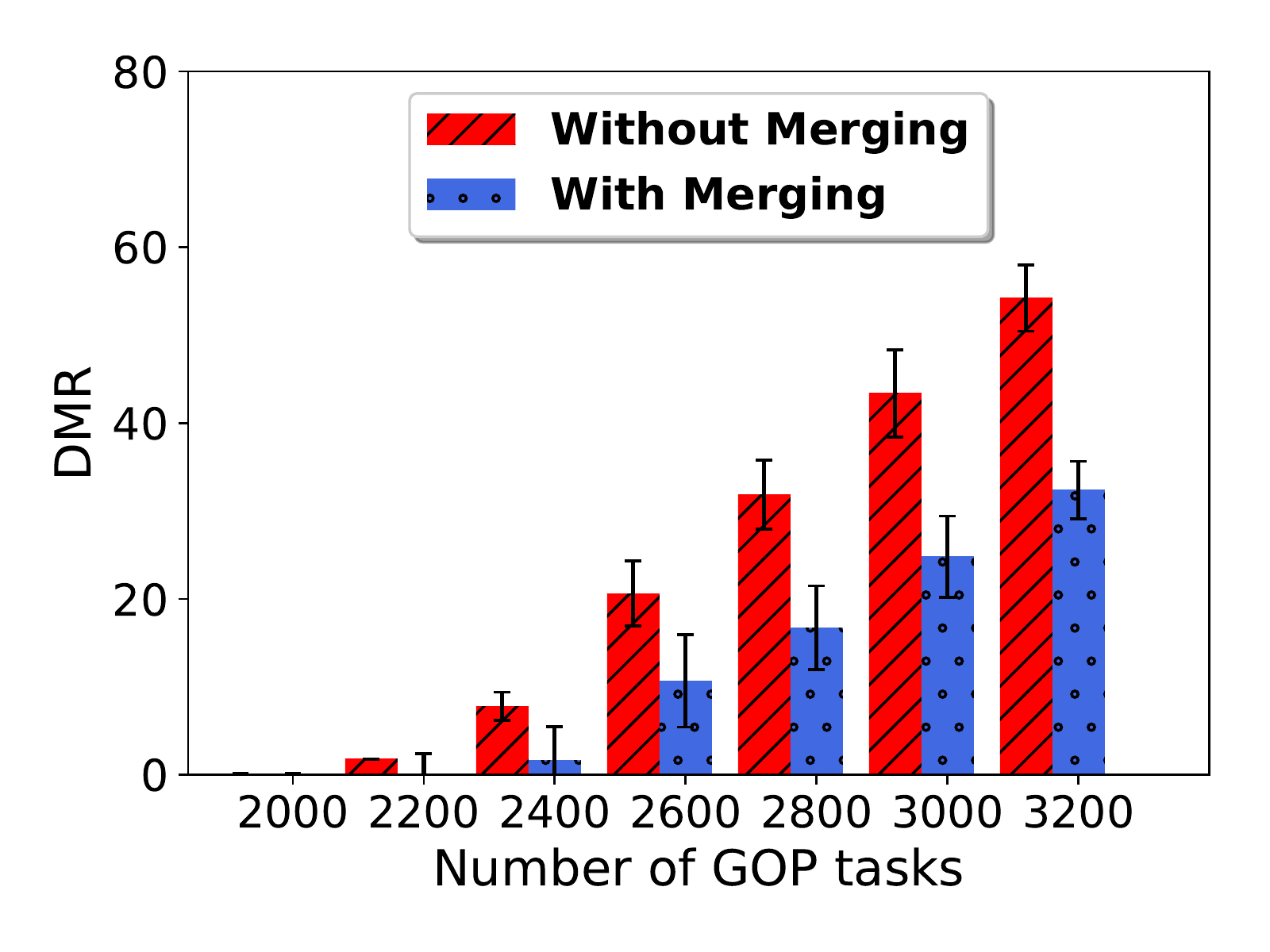} }}%
    
    \caption{\small{Comparing the deadline miss-rate (DMR) under varying number of GOP tasks (horizontal axes) in two scenarios: without task merging, and with  task merging. Subfigures (a), (b), and (c) show the DMR under FCFS, EDF, and MU queuing polices.}}    
    \label{fig:deadline}%
\end{figure*}

\paragraph*{Evaluating Deadline Miss Rate (DMR):}
In this experiment, our goal is to evaluate viewers' QoE. For that purpose, we measure the deadline miss rate resulted from no merging versus merging tasks under various oversubscribed levels and with different scheduling policies. 
As shown in Figure~\ref{fig:deadline}, we observe that task aggregation significantly reduces deadline miss rate in all scheduling policies. We can see that the improvement in deadline miss rate of FCFS is less than EDF and MU scheduling policies. This is because FCFS by nature causes a larger average waiting time and does not schedule tasks by considering their deadline. Therefore, task merging mechanism performance, when combined with FCFS, is lower than other scheduling polices. 

Comparing the results shown in Figure~\ref{fig:exetime} with those in Figure~\ref{fig:deadline} reveals that the difference in deadline miss rate is more dramatic than the makespan time. This is due to the fact that even small reduction in task completion time can cause the merged tasks meet their deadlines instead of missing that. We can conclude that the impact of task aggregation mechanism on viewers' QoE can become more remarkable when it is combined with efficient scheduling policies.

\section{Related works}\label{sec:relwk}
Software-based computational reuse has been extensively researched and used. However, not many systems can merge and reuse tasks before tasks are actually executed and many of them tie very closely to one specific application. Below are some notable works in this area.

Popa \emph{et al.}~\cite{DryadInc} presented modules to identify identical and similar tasks to cache partial results and reuse them on incremental computation specifically on Dryad platform context. They proposed two solutions: One solution automatically caches computational results. Another solution merges tasks based on programmer's defined merge function. Their first solution is a caching system while their second solution is similar to our work, but more specific to Dryad platform which does not have deadline and QoE to consider.

Paulo and Pereira \emph{et al.}~\cite{paulo2014distributed}
developed a system to perform deduplication of high throughput data using $Bloom filters$. $Bloom filters$, while fast, have chances of giving false positive hash checking. Therefore they achieve lower overhead data duplicate detection than hash table approach we use, at the price of compromised accuracy.
\section{Conclusion and Future works}\label{sec:conclsn}
In this paper, we improve efficiency of the system in oversubscribed condition by merging arriving tasks with other (exact or similar) tasks. 
We dealt with two challenges: First, how to identify identical and similar tasks in an efficient manner? Second, how to perform merging without violating the deadline of other tasks in the system?   
To address the first challenge, we identified three main levels of similarity that tasks can be merged. Then, we developed a method to detect different levels of task similarity within a constant time complexity. To address the second challenge, we developed a method that determines impact of merging and only perform merge operations if other tasks' deadline are not affected. Experimental results demonstrate that the proposed system can reduce the overall execution time of tasks by more than 14\%, hence, cloud VMs can be deployed for a shorter time. This benefit comes with improving QoE of the users.
Although we implemented this system in the context of video streaming, the concept can be applied to other domains as long as we can define similarity levels in those domains.
In the future, we plan to extend this work by exploring the impact of marginally compromising QoE, in favor of a remarkable cost-saving on the cloud resources.




\section*{Acknowledgments}
This research was supported by the Louisiana Board of Regents under grant number LEQSF(2016-19)-RD-A-25. 

\bibliographystyle{abbrv}
\bibliography{references}
\end{document}